\documentclass[%
reprint,
amsmath,amssymb,
aps,
]{revtex4-2}

\usepackage{graphicx}% Include figure files
\usepackage{dcolumn}% Align table columns on decimal point
\usepackage{bm}% bold math
\usepackage[colorlinks=true,citecolor=blue,urlcolor=black,linkcolor=black,filecolor=cyan]{hyperref}

\usepackage[capitalise]{cleveref}
\graphicspath{{figs/}}

\begin{document}

\title{Voltage Hysteresis of Silicon Nanoparticles: Chemo-Mechanical Particle-SEI Model}

\author{Lukas Köbbing}
\affiliation
{Institute of Engineering Thermodynamics, German Aerospace Center (DLR), Wilhelm-Runge-Straße 10, 89081 Ulm, Germany}
\affiliation
{Helmholtz Institute Ulm (HIU), Helmholtzstraße 11, 89081 Ulm, Germany}

\author{Arnulf Latz}
\author{Birger Horstmann}
\email{birger.horstmann@dlr.de}
\affiliation{Institute of Engineering Thermodynamics, German Aerospace Center (DLR), Wilhelm-Runge-Straße 10, 89081 Ulm, Germany}
\affiliation
{Helmholtz Institute Ulm (HIU), Helmholtzstraße 11, 89081 Ulm, Germany}
\affiliation{Institute of Electrochemistry, Ulm University, Albert-Einstein-Allee 47, 89081 Ulm, Germany}

\begin{abstract}
  Silicon is a promising anode material for next-generation lithium-ion batteries. However, the volume change and the voltage hysteresis during lithiation and delithiation are two substantial drawbacks to their lifetime and performance. We investigate the reason for the voltage hysteresis in amorphous silicon nanoparticles covered by a solid-electrolyte interphase (SEI). Concentration gradients inside the nanoscale silicon can not produce the massive stresses necessary to cause the reported voltage hysteresis. Our chemo-mechanical model shows that plastic deformation of the stiff, inorganic SEI during lithiation and delithiation reproduces the observed silicon open-circuit voltage hysteresis. Additionally, the viscous behavior of the SEI explains the difference between the voltage hysteresis observed at low currents and after relaxation. We conclude that the visco-elastoplastic behavior of the SEI is the origin of the voltage hysteresis in silicon nanoparticle anodes. Thus, consideration of the SEI mechanics is crucial for further improvements.
\end{abstract}

\maketitle

\section{Introduction}
The improvement of lithium-ion batteries is essential for facing the pressing global challenges. Due to their high theoretical capacity, pure silicon anodes are a promising candidate for the next generation of lithium-ion batteries \cite{Zuo2017, Kim2021, Zhang2021, Sun2022, Lin2023}. Therefore, research makes an effort to investigate this advantageous and abundant material.

Silicon confronts research with two challenging features on the way to commercial silicon anodes. The first is the significant expansion and constriction with a volume change of up to $300\%$ during lithiation and delithiation, respectively. The deformation leads to the cracking of large silicon particles and the formation of networks with nanometer-sized silicon particles \cite{Wetjen2018}. However, no cracking occurs below a critical particle diameter of about $150\,\mathrm{nm}$ \cite{Liu2012}. Due to the superior mechanical stability, recent research concentrates on investigating nano-structured silicon anodes \cite{Chakraborty2015, Stein2016, Yang2020, Xiao2022, Zhang2022}. Additionally, experiments observe a volume hysteresis, i.e. a difference in the silicon particle or anode volume between lithiation and delithiation \cite{Yoon2016, Duay2016, Jerliu2018, Dong2019}.

The second major issue is the voltage hysteresis of silicon anodes observed in experiments for various silicon structures, namely thin films \cite{Baggetto2009, Sethuraman2010a, Sethuraman2010b, Sethuraman2013, Verbrugge2015, Yoon2016, Lu2016, Baker2017}, nanowires \cite{Chan2008, Wu2012}, and nanoparticles \cite{Chandrasekaran2011, Bernard2019, Pan2019, Pan2020}. The voltage hysteresis considerably reduces energy efficiency and leads to detrimental heat generation during cycling \cite{McDowell2013}. 
Thus, the hysteresis phenomenon is a very important challenge for the commercialization of silicon-anode lithium-ion batteries.
Most literature reports the hysteresis of pseudo-open-circuit voltages at currents smaller than $\mathrm{C}/10$. However, even for the open-circuit voltage measured by the galvanostatic intermittent titration technique (GITT) after relaxation periods, a reduced, but still significant voltage hysteresis remains \cite{Sethuraman2013, Pan2019, Pan2020}. We emphasize that the measurements reveal a clear difference in the size of the voltage hysteresis observed for small currents and after relaxation periods.

Phase transformations can explain a voltage hysteresis for crystalline silicon or the first cycle of amorphous silicon anodes \cite{Yang2012, Cui2013, Zhao2019, Jiang2020}. However, also amorphous silicon anodes show a voltage hysteresis after the first cycle. In this case, the literature commonly considers plastic flow of silicon as the reason for the hysteresis phenomenon. In thin-film silicon electrodes, massive stresses arise naturally due to the restricted expansion in the in-plane direction \cite{Bower2011, DiLeo2015, Lu2016}. For micron-sized silicon particles or high applied currents, the slow diffusion in silicon can lead to concentration gradients during lithiation and delithiation, inducing a voltage hysteresis \cite{Chandrasekaran2010, Zhao2011, Cui2012}. Nevertheless, the reason for substantial stresses inside amorphous silicon nanoparticles during slow lithiation and delithiation remains unclear.

Similarly to other anode materials, the solid-electrolyte interphase (SEI) naturally covers silicon particles. The SEI protects the anode from the electrolyte \cite{Horstmann2021, Horstmann2019} and grows via electron transport from the anode through the SEI \cite{Kolzenberg2020, Koebbing2023}. Additionally, between the silicon particle and the SEI, there is a native silicon oxide layer influencing the interface between particle and SEI \cite{Cao2019}. Research invests much effort into characterizing and improving the SEI on silicon anodes \cite{Pinson2013, Nie2013, Wetjen2017, Zhang2019, Kim2021sei, Zhang2021interplay, Dopilka2023}. Due to the considerable volume changes of silicon anodes, the plasticity and cracking of the SEI on silicon deserve particular interest \cite{Kumar2017, Tanaka2018, Guo2020, Jin2021, Kolzenberg2021}.

As the SEI deforms due to the chemical expansion and shrinkage of the particle, considerable strains and stresses occur inside the SEI. Certain studies even claim that the compressive stress of the SEI acting on the particle might be beneficial to avoid particle cracking \cite{Li2019, Chen2020, Sun2022}. Further, we note that a carbon coating on top of the silicon particle reduces the observed voltage hysteresis of the silicon anode \cite{Bernard2019}. The reason might be a different SEI composition on carbon and silicon. In conclusion, we propose that the composition and stress of the SEI have a crucial impact on the lithiation behavior and, particularly, on the voltage hysteresis of silicon particles. 

In this paper, we consider a visco-elastoplastic SEI model based on non-equilibrium thermodynamics \cite{Schammer2021}. We evaluate the influence of the stress generated by the SEI on the lithiation and delithiation behavior of an amorphous silicon nanoparticle. Our model can faithfully reproduce the broadly recognized but empirical Plett model. Furthermore, it allows to investigate the reason for the difference between the voltage hysteresis for small currents and GITT data points. Finally, we discuss the influence of the mechanical parameters of the SEI and the silicon nanoparticle on the silicon voltage hysteresis.

\section{Theory}

\begin{figure*}[htbp]
	% Figure #?
	% width = 2 columns
	\centering
	\includegraphics[width=0.9\textwidth]{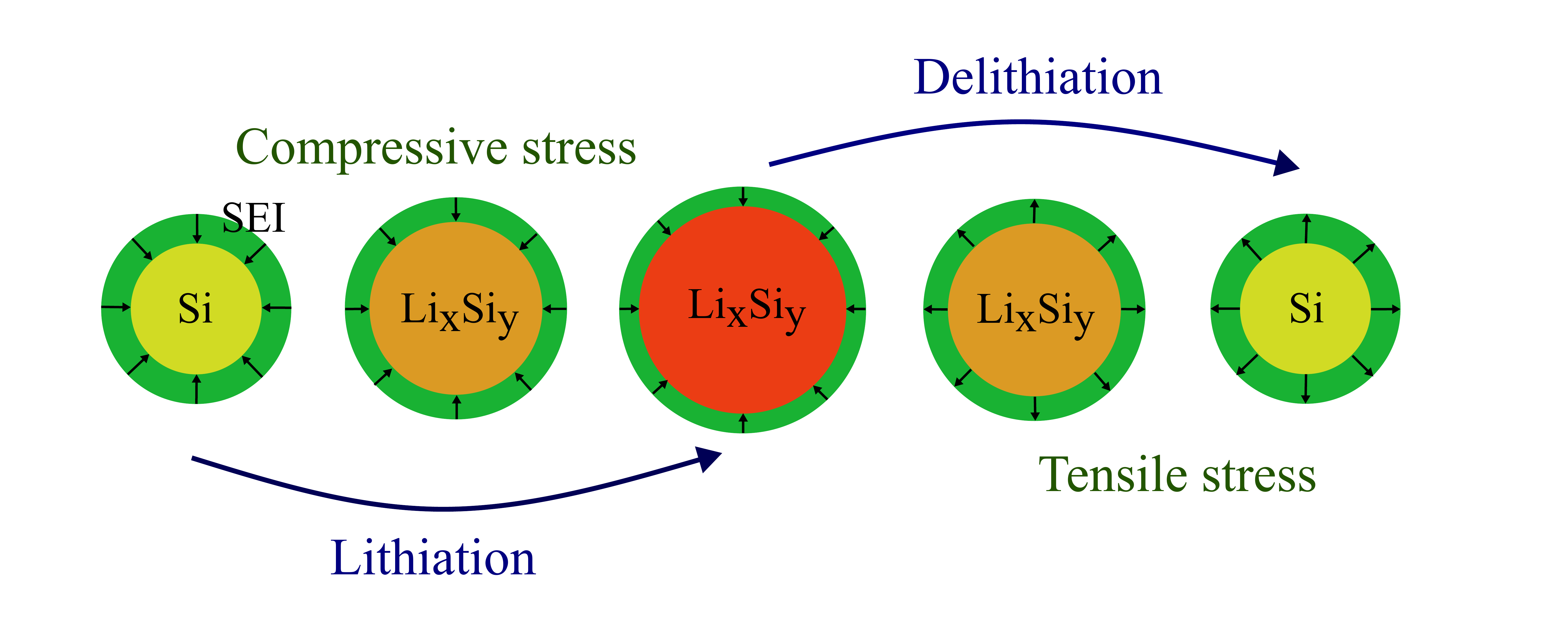}
	\caption{Scheme of the radial stress inside the SEI during lithiation and delithiation of a silicon nanoparticle.}
	\label{fig:scheme-particle-sei}
\end{figure*}

When the silicon anode particle expands and contracts, distinct strains emerge inside the SEI as illustrated in \cref{fig:scheme-particle-sei}. Elastic deformations lead to significant stresses in tangential and radial directions. The radial stress component impacts the stress inside the particle, which crucially influences the silicon potential. We discuss a chemo-mechanical silicon particle model and a visco-elastoplastic SEI model.

\subsection{Silicon particle model}
Our silicon particle model is based on the work of Kolzenberg et al. complemented by plasticity \cite{Kolzenberg2021, Castelli2021}. During lithiation and delithiation, the silicon particle experiences significant expansion and shrinkage, respectively. The deformation tensor $\mathbf{F} = \partial \vec{x}/\partial \vec{X}_0$ describes the transformation from the undeformed Lagrangian frame to the deformed Eulerian frame \cite{Holzapfel2000}. We use the large deformation approach to describe the deformation inside the silicon particle \cite{Gurtin2010}. Therefore, the total deformation $\mathbf{F}$ consisting of a chemical deformation $\mathbf{F}_\mathrm{ch}$, an elastic deformation $\mathbf{F}_\mathrm{el}$, and a plastic deformation $\mathbf{F}_\mathrm{pl}$ reads
\begin{equation}
	\mathbf{F} = \mathbf{F}_\mathrm{pl} \mathbf{F}_\mathrm{el} \mathbf{F}_\mathrm{ch} .
\end{equation}
The chemical deformation $\mathbf{F}_\mathrm{ch}$ is described by the concentration of lithium $c_\mathrm{Li,0}$ inside the silicon particle in the Lagrangian frame according to
\begin{equation}
	\mathbf{F}_\mathrm{ch} = \lambda_\mathrm{ch} \mathbf{Id} = (1+v_\mathrm{Li} c_\mathrm{Li,0})^{1/3} \mathbf{Id},
\end{equation}
where $v_\mathrm{Li}$ is the molar volume of lithium inside silicon.

The lithiation behavior and the mechanical deformations are derived in our model based on non-equilibrium thermodynamics. Our model builds up on the free energy according to the first law of thermodynamics. In section SI in the supporting information, we discuss non-negativity of the entropy production $\mathcal{R} \geq 0$ in our model stated in Eq. (S12), which means accordance with the second law.

The free energy density $\rho_0 \varphi$ of the silicon anode consists of a chemical contribution depending on the lithium concentration and a mechanical contribution due to elastic deformation. The free energy of the silicon anode reads
\begin{equation}
\begin{aligned}
	\rho_{0} \varphi = &-\int_0^{c_{\mathrm{Li}}} F U_0\left(c_{\mathrm{Li}}^{\prime}\right) \mathrm{d} c_{\mathrm{Li}}^{\prime}\\
	 &+ \frac{1}{2}\left(\lambda_\mathrm{Si}\left(\operatorname{tr}\left(\mathbf{E}_{\mathrm{el}}\right)\right)^2+2 G_\mathrm{Si} \operatorname{tr}\left(\mathbf{E}_{\mathrm{el}}^2\right)\right)
    \label{eq:free-energy}
\end{aligned}
\end{equation}
with the Faraday constant $F$ and the ideal open-circuit voltage of silicon $U_0$. The elastic strain tensor $\mathbf{E}_\mathrm{el}$ in \cref{eq:free-energy} is defined by
\begin{equation}
\mathbf{E}_\mathrm{el} = \frac{1}{2} \left(\mathbf{F}_\mathrm{el}^\mathrm{T}\mathbf{F}_\mathrm{el}^{\phantom{\mathrm{T}}}-\mathbf{Id}\right).
\end{equation}
The first and second Lamé constants read $\lambda_\mathrm{Si} =2 G_\mathrm{Si} \nu_\mathrm{Si} /\left(1-2 \nu_\mathrm{Si}\right)$ and $G_\mathrm{Si} = E_\mathrm{Si}/2(1+\nu_\mathrm{Si})$, where $\nu_\mathrm{Si}$ is the Poisson ratio of silicon. The elastic deformation $\mathbf{F}_\mathrm{el}$ depends on the Piola-Kirchhoff stress $\mathbf{P}$ inside the particle determined by the constitutive equation
\begin{align}
	\mathbf{P} &= 2\mathbf{F}\frac{\partial \rho_0 \varphi}{\partial \mathbf{C}}\label{eq:stress-particle}\\
 &= \lambda_\mathrm{ch}^{-2} \mathbf{F} \mathbf{F}_\mathrm{pl}^{-\mathrm{T}} \mathbf{F}_\mathrm{pl}^{-1} \left(\lambda_\mathrm{Si} \operatorname{tr}(\mathbf{E}_\mathrm{el})\mathbf{Id} + 2G_\mathrm{Si} \mathbf{E}_\mathrm{el}\right),\nonumber
\end{align}
where $\mathbf{C}$ is the right Cauchy-Green tensor $\mathbf{C} = \mathbf{F}^\mathrm{T} \mathbf{F}$. The
Piola-Kirchhoff stress is related to the Cauchy stress $\sigma$ as $\mathbf{P} = J \sigma \mathbf{F}^{-\mathrm{T}}$ with $J=\operatorname{det} \mathbf{F}$.
The Piola-Kirchhoff stress $\mathbf{P}$ defined in \cref{eq:stress-particle} has to fulfill the momentum balance in the Lagrangian frame
\begin{equation}
	0 =\nabla_{0} \cdot \mathbf{P}.
    \label{eq:nabla-stress}
\end{equation}
In addition to the elastic deformation, the particle deforms plastically when reaching the von Mises yield criterion
\begin{equation}
	f = \frac{\frac{3}{2} |\mathbf{M}^\mathrm{dev}|^2}{\sigma_\mathrm{Y,Si}^2} - 1 \leq 0,
\end{equation}
where $\mathbf{M}^\mathrm{dev} = \mathbf{M} - 1/3\operatorname{tr} \mathbf{M}$ is the deviatoric part of the Mandel stress $\mathbf{M} = \mathbf{F}_\mathrm{rev}^\mathrm{T} \sigma \mathbf{F}_\mathrm{rev}^{-\mathrm{T}}$ with the reversible part of the deformation $\mathbf{F}_\mathrm{rev}=\mathbf{F}_\mathrm{el}\mathbf{F}_\mathrm{ch}$. The yield stress $\sigma_\mathrm{Y}$ depends on the material and reveals how much stress the particle can withstand with reversible deformations. The plastic flow $\mathbf{L}_\mathrm{pl} = \dot{\mathbf{F}}_\mathrm{pl} \mathbf{F}_\mathrm{pl}^{-1}$ is calculated by
\begin{equation}
	\mathbf{L}_\mathrm{pl} = \phi \frac{\partial f}{\partial \mathbf{M}},
    \label{eq:plastic-flow}
\end{equation}
where the plastic multiplier $\phi$ is determined from the consistency condition $\dot{f} = 0$.

The defining equation for the lithiation and delithiation of a silicon particle reads
\begin{equation}
	\dot{c}_{\mathrm{Li}, 0} =-\nabla_{0} \cdot \vec{N}_{\mathrm{Li}, 0}
    \label{eq:dot-concentration}
\end{equation}
with the lithium flux $\vec{N}_{\mathrm{Li}, 0} = -L \nabla_0 \mu_\mathrm{Li}$ and the electro-chemo-mechanical potential $\mu_\mathrm{Li}$. The mobility is determined by $L=D_\mathrm{Li} \left(\partial \mu_\mathrm{Li}/\partial c_\mathrm{Li,0}\right)^{-1}$ with $D_\mathrm{Li}$ the diffusion coefficient of lithium in silicon. At the boundary of the particle $r=R$, the lithium flux $\vec{N}_{\mathrm{Li}, 0}(R)$ is determined by the applied (de)lithiation rate.

The lithium concentration and the stress inside the particle influence the electro-chemo-mechanical potential of lithium according to the constitutive equation
\begin{equation}
	\mu_\mathrm{Li} = \frac{\partial \rho_0 \varphi}{\partial c_\mathrm{Li,0}} = - FU_0 - \frac{v_\mathrm{Li}}{3 \lambda_\mathrm{ch}^3} \mathbf{P} : \mathbf{F}.
	\label{eq:chemical-potential}
\end{equation}
The chemical potential at the outer border of the anode determines the measurable silicon open-circuit voltage (OCV) by $U=-\mu_\mathrm{Li}/F$. Thus, the OCV reads
\begin{equation}
	U = U_0 + \frac{v_\mathrm{Li}}{3 F \lambda_\mathrm{ch}^3} \mathbf{P} : \mathbf{F}.
	\label{eq:ocv}
\end{equation}
We determine the ideal open-circuit voltage $U_0$ as the mean value of the open-circuit voltages measured via GITT during lithiation and delithiation in Ref. \cite{Pan2019}. 
However, stress inside the particle crucially affects the open-circuit voltage during lithiation and delithiation.

\subsection{SEI model}
Inside the SEI, there is no chemical deformation, as no lithiation of the SEI is possible. Therefore, the SEI only deforms elastically and plastically
\begin{equation}
	\mathbf{F}_\mathrm{SEI} = \mathbf{F}_\mathrm{SEI,pl} \mathbf{F}_\mathrm{SEI,el}.
\end{equation}
In our visco-elastoplastic SEI model, elastoplastic deformations and viscous flow contribute to the total stress. The stress due to elastoplastic deformations is defined similarly to \cref{eq:stress-particle} as
\begin{align}
	&\mathbf{P}_\mathrm{SEI,el} = 2\mathbf{F}_\mathrm{SEI}\frac{\partial \rho_\mathrm{0,SEI} \varphi_\mathrm{SEI}}{\partial \mathbf{C}_\mathrm{SEI}}\\ &= \mathbf{F}_\mathrm{SEI} \mathbf{F}_\mathrm{SEI,pl}^{-\mathrm{T}}  \mathbf{F}_\mathrm{SEI,pl}^{-1} \left(\lambda_\mathrm{SEI} \operatorname{tr}(\mathbf{E}_\mathrm{SEI,el})\mathbf{Id} + 2G_\mathrm{SEI}\mathbf{E}_\mathrm{SEI,el}\right)\nonumber
\end{align}
with the strain tensor $\mathbf{E}_\mathrm{SEI,el}$, the right Cauchy-Green tensor $\mathbf{C}_\mathrm{SEI}$, and the Lamé constants $\lambda_\mathrm{SEI}$ and $G_\mathrm{SEI}$ defined analog to the particle model. The free energy density of the SEI is defined according to merely the mechanical contribution in \cref{eq:free-energy}.

Plastic flow can occur inside the SEI similarly to the particle when reaching the von Mises yield criterion
\begin{equation}
	f_\mathrm{SEI} = \frac{\frac{3}{2} |\mathbf{M}_\mathrm{SEI,el}^\mathrm{dev}|^2}{\sigma_\mathrm{Y,SEI}^2} - 1 \leq 0
\end{equation}
leading to the plastic flow
\begin{equation}
	\mathbf{L}_\mathrm{SEI,pl} = \phi_\mathrm{SEI} \frac{\partial f_\mathrm{SEI}}{\partial \mathbf{M}_\mathrm{SEI,el}}.
    \label{eq:plastic-flow-sei}
\end{equation}
The consistency condition $\dot{f}_\mathrm{SEI}=0$ determines again the value of the plastic multiplier $\phi_\mathrm{SEI}$. We emphasize that plastic flow in the SEI is driven by the deviatoric part of the elastoplastic stress contribution $\mathbf{M}_\mathrm{SEI,el}^\mathrm{dev}$.

In addition to the elastoplastic model discussed in Ref. \cite{Kolzenberg2021}, we consider stress generated by the viscous flow of the SEI. The viscous Cauchy stress is defined as
\begin{equation}
	\mathbf{\sigma}_\mathrm{SEI,visc} = \eta_\mathrm{SEI} \dot{\mathbf{E}}_\mathrm{SEI}
    \label{eq:sigma-viscous}
\end{equation}
with the viscosity of the SEI $\eta_\mathrm{SEI}$. Radial symmetry in our system ensures $\dot{\mathbf{E}}_\mathrm{SEI} = \mathbf{F}_\mathrm{SEI}^\mathrm{T} \dot{\mathbf{F}}_\mathrm{SEI}$.
Therefore, the viscous Piola-Kirchhoff stress corresponding to \cref{eq:sigma-viscous} reads
\begin{equation}
	\mathbf{P}_\mathrm{SEI,visc} = J_\mathrm{SEI} \eta_\mathrm{SEI} \dot{\mathbf{F}}_\mathrm{SEI}
\end{equation}
with $J_\mathrm{SEI} = \operatorname{det} \mathbf{F}_\mathrm{SEI}$.

We describe the SEI as a power-law shear thinning material. Therefore, the viscosity decreases with increasing strain rate
\begin{equation}
	\eta_\mathrm{SEI} \left(\dot{\mathbf{E}}_\mathrm{SEI}\right) = \eta_\mathrm{SEI,0} \dot{\mathbf{E}}_\mathrm{SEI}^{n-1},
\end{equation}
where $\eta_\mathrm{SEI,0}$ is a constant value and  $n<1$ is the shear-thinning exponent. 

In our model, we consider the viscous behavior in parallel to the elastoplastic model. Thus, the sum of the elastic and viscous contributions describes the total Piola-Kirchhoff stress as
\begin{equation}
	\mathbf{P}_\mathrm{SEI} = \mathbf{P}_\mathrm{SEI,el} + \mathbf{P}_\mathrm{SEI,visc}.
\end{equation}
Analog to the particle model, the momentum balance
\begin{equation}
	0 = \nabla_{0} \cdot \mathbf{P}_\mathrm{SEI}
 \label{eq:nabla-stress-sei}
\end{equation}
determines the stress inside the SEI.

Finally, we emphasize that our model describes the SEI mechanics on a continuum level. The visco-elastoplastic behavior is not a single-crystal property but results from subsequent partial SEI fracture and healing \cite{Kolzenberg2021}. This approach is valid for small currents, where the SEI fractures only partially and healing is fast compared to the lithiation timescale.

\subsection{Particle-SEI interface}
At the interface between the anode particle and SEI, we assume a perfect sticking of the SEI on top of the silicon particle. Therefore, the stresses inside the particle and SEI interact at the interface. Precisely, the radial component of the Cauchy stress has to be equal in particle and SEI at the boundary. Due to the radial symmetry, this also implies the equivalence of the radial part of the Piola-Kirchhoff stress at the interface
\begin{equation}
	\mathbf{P}_\mathrm{rr} \big\rvert_{r=R} = \mathbf{P}_\mathrm{SEI,rr} \big\rvert_{r=R} .
\end{equation}

Inside the particle, the chemical contribution dominates the total deformation. In contrast, the SEI deforms only elastically and plastically. Thus, mechanical deformations of the SEI have to afford the large deformation implied by the particle. These deformations result in significant strains and stresses inside the SEI. Consequently, the stress inside the SEI can substantially affect the stress and lithiation behavior of the silicon particle.

\subsection{Material parameters}
\label{sec:material-parameters}

The literature states values of Young's modulus of the SEI $E_\mathrm{SEI}$ between only a few hundreds of $\mathrm{MPa}$ and values greater than $100\,\mathrm{GPa}$. For a better overview, we divide the reported values into the categories:
soft with $E_\mathrm{SEI}<1\,\mathrm{GPa}$ \cite{Zheng2014, Kuznetsov2015, Yoon2020, Gu2023, Chen2023}, 
medium with $1\,\mathrm{GPa} \leq E_\mathrm{SEI} < 10\,\mathrm{GPa}$ \cite{Zheng2014, Shin2015, Shkrob2015, He2017, Kamikawa2020, Liu2022, Gu2023},
stiff with $10\,\mathrm{GPa} \leq E_\mathrm{SEI} < 100\,\mathrm{GPa}$ \cite{Zhang2015, Shin2015, Shkrob2015, He2017, Tanaka2018, Zhang2020, Chai2021},
and very stiff SEI with $E_\mathrm{SEI} \geq 100\,\mathrm{GPa}$ \cite{Shin2015, Chai2021}.
In general, references report lower values of Young's modulus for thicker SEI layers. In contrast, the initial SEI or the inner, inorganic SEI layer reveals higher values of Young's modulus. Due to the small scale, the measurement values of the inner SEI might underestimate the corresponding Young's modulus, as the measured point might be in the transition zone between the inner and the outer SEI. Further, the literature reports a size effect of Young's modulus on the nanoscale for different materials. Structures on the nanometer level typically reveal a higher Young's modulus compared to bulk values \cite{Chen2006, Mathur2007, Agrawal2008, He2008, Chen2015}. Therefore, we use $E_\mathrm{SEI} = 100\,\mathrm{GPa}$ for the SEI in our studies.

Experiments attempt to determine the viscosity of the SEI but face severe difficulties due to the small length scale of interest. Recently, research has advanced in estimating the viscosity value of the outer, organic SEI layer \cite{Chai2021, Chen2023}. However, the inner, inorganic SEI layer is known to be much stiffer than the outer layer. We assume this coincides with a considerably higher viscosity of the inner layer. Consequently, reasonable values of the viscosity of the inner SEI are between $\eta = 10^7\,\mathrm{Pa\,s}$ measured for pitch, a highly viscous polymer \cite{Edgeworth1984}, and $\eta = 10^{15}\,\mathrm{Pa\,s}$ measured for silicon oxide \cite{Sutardja1989, Senez1994, Ojovan2008}. We estimate the shear-thinning exponent to be $n=0.15$ from Ref. \cite{LePage2019}.

Measurements for Young's modulus of silicon show a wide range of values \cite{Allred2004, NasrEsfahani2019}. 
For amorphous macroscopic silicon, literature reports values around $90\,\mathrm{GPa}$ \cite{Kluge1988, Shenoy2010} and $125\,\mathrm{GPa}$ \cite{Tan1972, Szabadi1998}. Crystalline silicon nanowires reveal values between $50\,\mathrm{GPa}$ and $250\,\mathrm{GPa}$ \cite{NasrEsfahani2019, Tang2012}.
At the nanoscale, we expect that the values of Young's modulus of amorphous and crystalline silicon will approach each other as crystalline phases inside nanoparticles gain more importance. In comparison to nanowires, nanoparticles may even show more pronounced surface effects leading to higher values of Young's modulus. Therefore, we use $E_\mathrm{Si} = 200\,\mathrm{GPa}$ in our simulations.

If not mentioned explicitly, we use the parameters presented in Table S1 in the supporting information.

\subsection{Computational details}

We implement our model in MATLAB using a finite difference approach. We solve the partial differential equations (\ref{eq:nabla-stress}), (\ref{eq:plastic-flow}), (\ref{eq:dot-concentration}),   (\ref{eq:plastic-flow-sei}), and (\ref{eq:nabla-stress-sei}) with the solver ode15i by discretizing the radial dimension. The principal variables inside the particle are the lithium concentration $c_\mathrm{Li,0}$, the deformed radius $r$, and the radial component of the plastic deformation $\mathbf{F}_\mathrm{pl,rr}$ of each element. Inside the SEI domain, the principal variables are the deformed radius $r_\mathrm{SEI}$ and the radial component of the plastic deformation $\mathbf{F}_\mathrm{SEI,pl,rr}$ of each SEI element.

\section{Results and Discussion}
\label{sec:results}

Based on our chemo-mechanical model, we evaluate the influence of the mechanics of the silicon particle and its covering interphase on the voltage hysteresis of silicon anodes in this section. Although we term the covering layer SEI throughout this paper, our model also describes a particle covered by a silicon oxide layer.

Without the impact of the SEI, we can create three different scenarios presented in the supporting information in sections SII to SIV reasoning the experimental voltage hysteresis.
Firstly, literature often considers silicon thin-film anodes, where the expansion and contraction are restricted to the normal direction of the film as illustrated in Figure S1. This leads to significant stresses and plastic flow, resulting in a voltage hysteresis shown in Figure S2.
Secondly, in a real multi-particle electrode, the expansion of a single particle is constrained. We model this situation with a simplified particle confined by a fixed wall (see Figure S3). As depicted in Figure S4, considerable stresses and a voltage hysteresis occur only during the first lithiation due to permanent plastic deformation.
Thirdly, massive stresses generated by concentration gradients can arise in silicon particles without constraints due to the slow diffusion process. However, S5 shows that no voltage hysteresis is visible for silicon nanoparticles cycled at $\mathrm{C}/20$. Only currents as large as $1\mathrm{C}$ create substantial concentration gradients and a voltage hysteresis in silicon nanoparticles without constraints. We emphasize that these concentration gradients vanish after relaxation and can not explain the open-circuit voltage hysteresis.

Consequently, plastic flow due to slow diffusion or constrained particles can not reproduce the voltage hysteresis observed in GITT measurements with a silicon nanoparticle anode. Thus, we investigate the influence of the SEI on the stress and potential inside the silicon particle.

\subsection{Open-circuit voltage hysteresis}

In our model, we consider the SEI as a visco-elastoplastic material. Inside the particle, a substantial amount of the total deformation is the chemical contribution due to the change in lithium concentration. In contrast, the SEI merely shows mechanical deformations. As the deformation of the SEI has to adjust to the deformation of the particle, considerable strains occur naturally inside the SEI. Assuming a high Young's modulus of the SEI, strains inside the SEI cause significant stresses. As discussed, the radial component of the stress inside the SEI at the particle-SEI interface determines the radial stress component inside the particle at the interface. For currents as small as $\mathrm{C}/20$, the radial stress component of the SEI is the main reason for the stress inside the particle. The particle stress distributes uniformly, reasoning vanishing deviatoric stress. Therefore, plastic flow will not occur inside the particle. Nevertheless, the substantial stresses inside the particle can generate a voltage hysteresis.

\begin{figure}[htbp]
	% Figure #?
	% width = 2 columns
	\centering
	\includegraphics[width=0.48\textwidth]{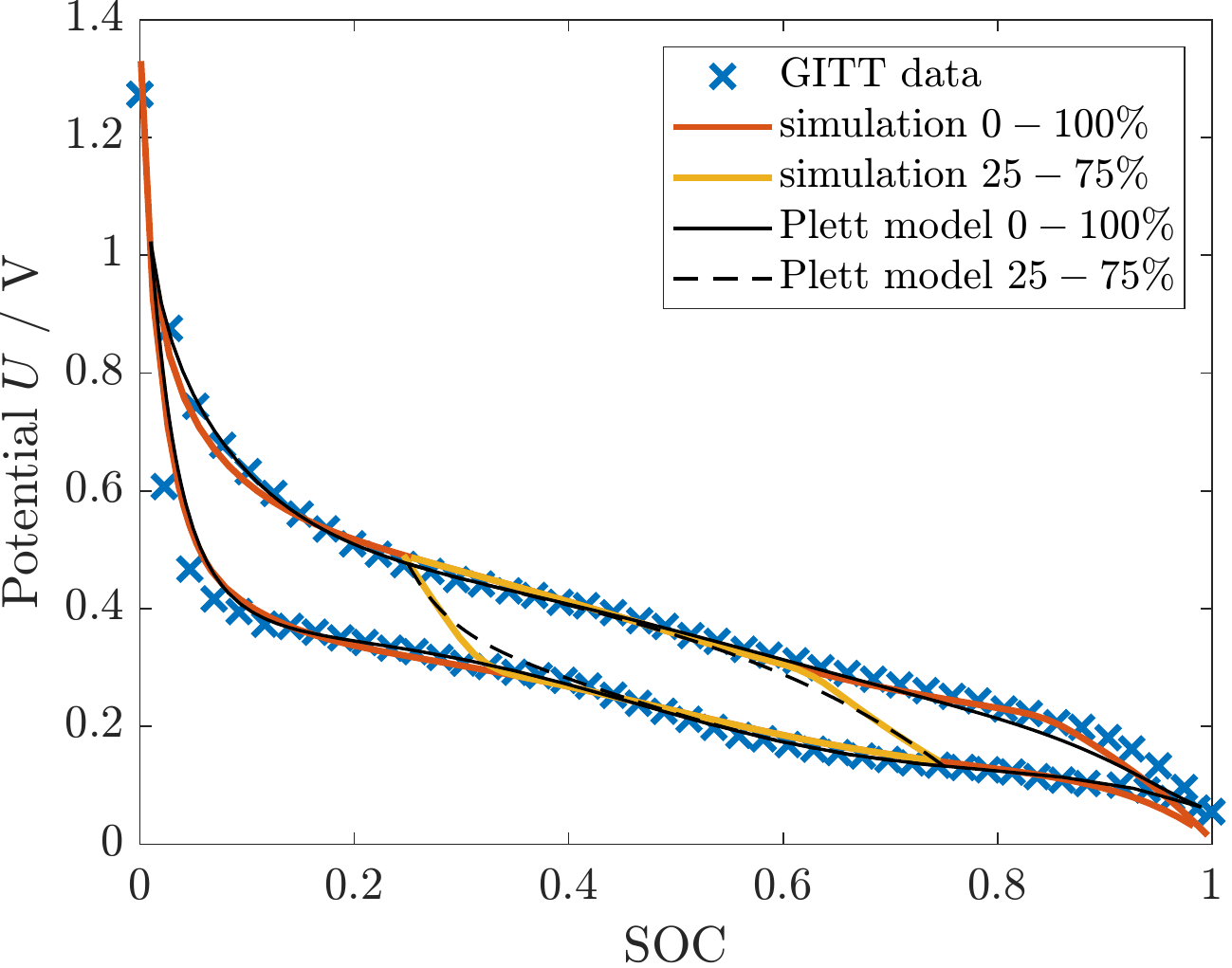}
	\caption{Open-circuit voltage hysteresis generated by a visco-elastoplastic SEI in comparison to GITT measurement \cite{Pan2019}. A complete cycle and a partial cycle are depicted for the simulation and the Plett model.}
	\label{fig:potential-hysteresis}
\end{figure}

In our simulation, we determine the open-circuit voltage after a relaxation period to reproduce GITT measurements. Figure \ref{fig:potential-hysteresis} depicts the open-circuit voltage for lithiation and delithiation simulated with our visco-elastoplastic SEI model depending on the state-of-charge (SOC). Notably, the simulation reveals a significant voltage hysteresis. Furthermore, our results agree well with the experimentally observed voltage hysteresis \cite{Pan2019}. Only for very low and very high SOC does the experiment show a steeper decrease and increase, respectively, compared to the simulation.
We attribute this to the use of the mean value of the lithiation and delithiation GITT voltages as the ideal open-circuit voltage curve. However, the situation is asymmetric at the endpoints of lithiation and delithiation. Before swapping the current direction, the stress stays constant but changes gradually after. Therefore, the ideal open-circuit voltage will differ from the mean value of the lithiation and delithiation open-circuit voltages at very low and very high SOC.

\vspace{0.2cm}
\vspace{0.2cm}

\begin{figure*}[htp]
	% Figure #?
	% width = 1 columns
	\centering
	\includegraphics[width=0.48\textwidth]{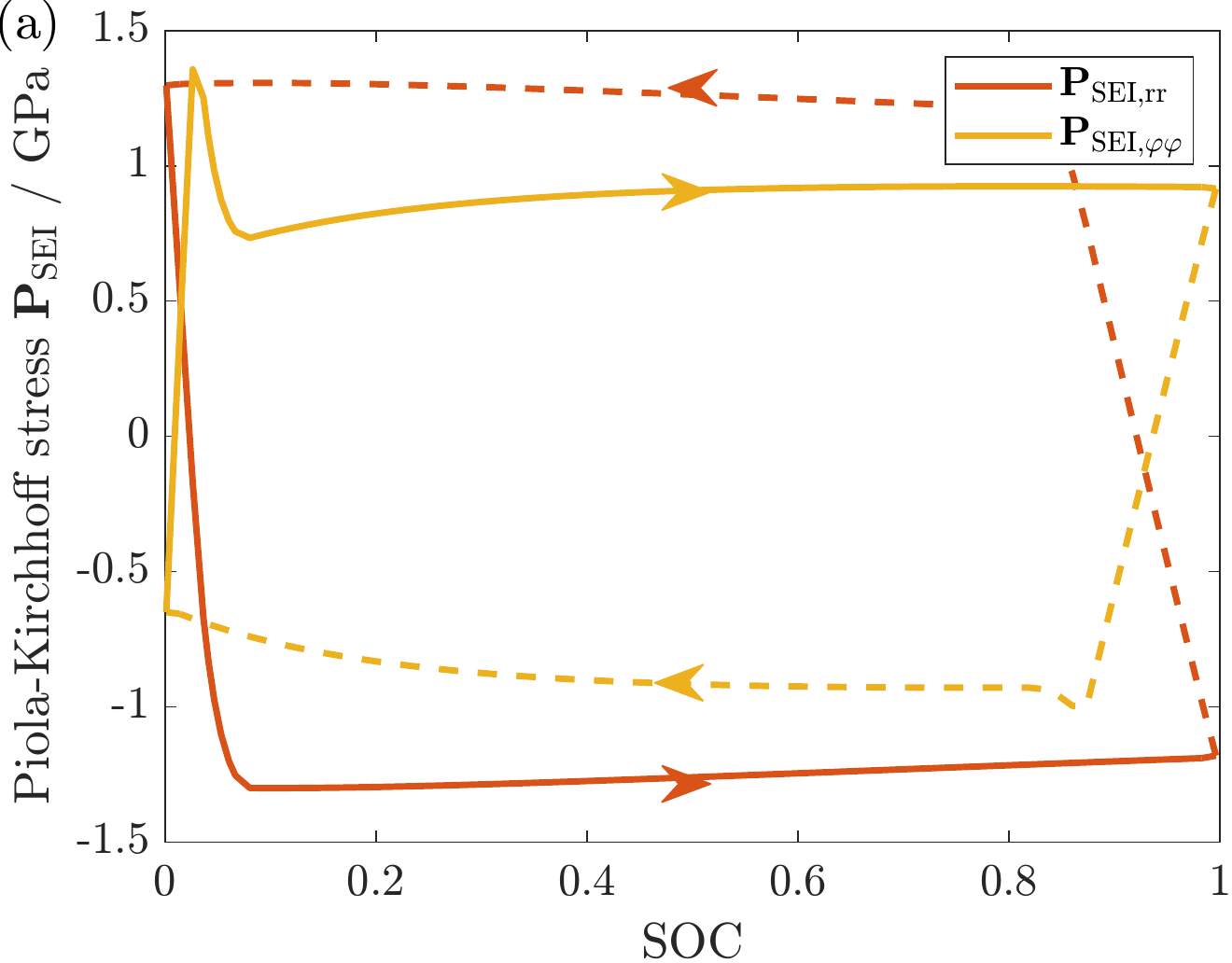}
	\hspace{0.5cm}
	\includegraphics[width=0.48\textwidth]{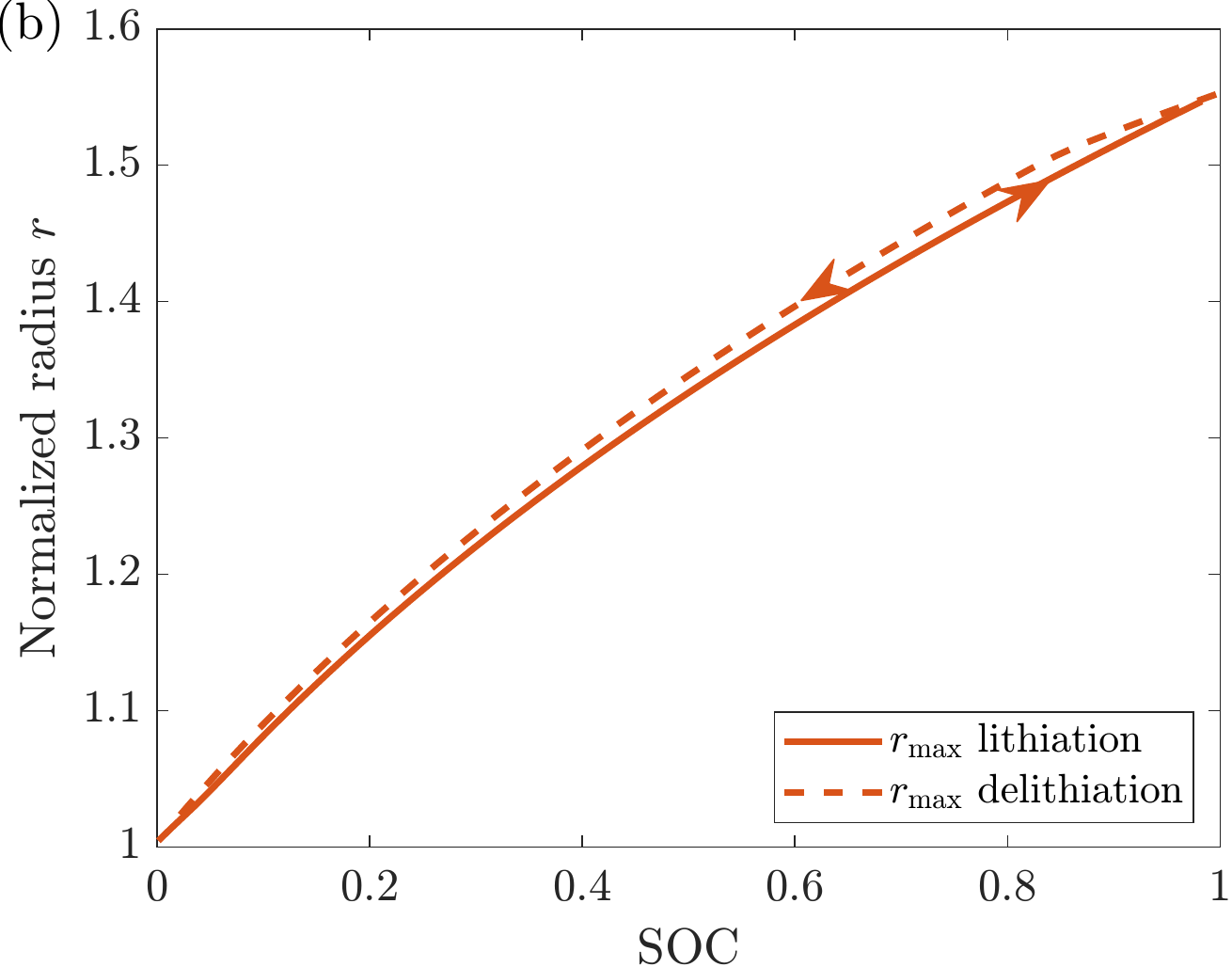}
	\caption{Stress and size effects of a visco-elastoplastic SEI. Lithiation is depicted with solid and delithiation with dashed lines. (a) Hysteresis of the stress components inside the SEI at the SEI-particle interface. (b) Hysteresis of the particle radius generated by the impact of the visco-elastoplastic SEI.}
	\label{fig:plots-viscous-GITT}
    %\vspace{-0.2cm}
\end{figure*}

The standard fitting model to describe transitions between the lithiation and delithiation voltages is the Plett model \cite{Plett2004, Graells2020, Wycisk2022}. We briefly present the equations describing the Plett model in the supporting information in section V. In \cref{fig:potential-hysteresis}, we depict our simulation of a complete cycle and a partial cycle between $25\%$ and $75\%$ in comparison to the behavior of the Plett model. The illustration reveals that our physical model is in reasonable accordance with the phenomenological Plett model.

In our model, the radial stress component of the SEI impacts the stress inside the particle and implicitly generates the voltage hysteresis of the silicon nanoparticle. In \cref{fig:plots-viscous-GITT}(a), we illustrate the stress components inside the SEI at the particle-SEI interface.
During lithiation, the particle expands, and the SEI deforms purely elastically until reaching the yield criterion. As the particle surface area increases, the SEI has to stretch in the tangential direction leading to tensile tangential stress (yellow). This tangential expansion leads to a constriction in the SEI thickness and compressive radial stress (red). Upon meeting the yield criterion, the SEI starts to flow plastically. The tangential stress shows a significant kink at the transition between elastic and plastic deformation. In the purely plastic regime, the stress components change only slightly. The radial component stays compressive, and the tangential tensile. Switching to delithiation, the SEI immediately leaves the plastic regime and deforms purely elastically again. The tensile tangential stress reduces, and compressive stress occurs. Simultaneously, compressive radial stress reduces, and tensile stress occurs. Reaching the plastic limit, the compressive tangential stress and the tensile radial stress stay almost constant until the end of the delithiation process. Eventually, the radial stress and the tangential stress of the SEI show a hysteresis behavior due to path-dependent plastic deformations.

The maximum stress in the system results from the yield condition, which determines the maximum deviatoric stress independent of the particle size. However, expressing \cref{eq:nabla-stress} in spherical coordinates indicates that the radial stress component increases with the ratio of SEI thickness and particle radius. Nevertheless, the particle size of interest is around the optimum size for silicon nanoparticles in terms of degradation found in Ref. \cite{Jin2021}. That means smaller particles with a higher specific surface area are covered by a thinner SEI compared to larger particles. We conclude that a change in particle size has only a minor influence on the size of the OCV hysteresis in the realm of interest.

As the stresses inside particle and SEI influence the particle deformations, the stress hysteresis induces a hysteresis phenomenon in the deformation tensor. This leads to a hysteresis behavior in the radius and volume of the particle. In \cref{fig:plots-viscous-GITT}(b), we depict the particle radius depending on the SOC, revealing a clear hysteresis effect. The particle radius shows larger values for delithiation than for lithiation at the same SOC. The simulation result matches qualitatively the volume hysteresis reported for silicon anodes \cite{Yoon2016, Duay2016, Jerliu2018, Dong2019}.

\subsection{Voltage hysteresis during cycling}

The literature states a significant difference between the voltage hysteresis of silicon anodes for small currents and the open-circuit voltage hysteresis measured with GITT. As discussed at the beginning of section \ref{sec:results}, we can not attribute this to concentration gradients inside the particle. Instead, we assign the potential difference to the viscous behavior of the SEI. As the literature provides no value for the viscosity of the inner SEI, we fit the viscosity to the experimental data.

\begin{figure*}[htp]
	% Figure #?
	% width = 2 columns
	\centering
	\includegraphics[width=0.48\textwidth]{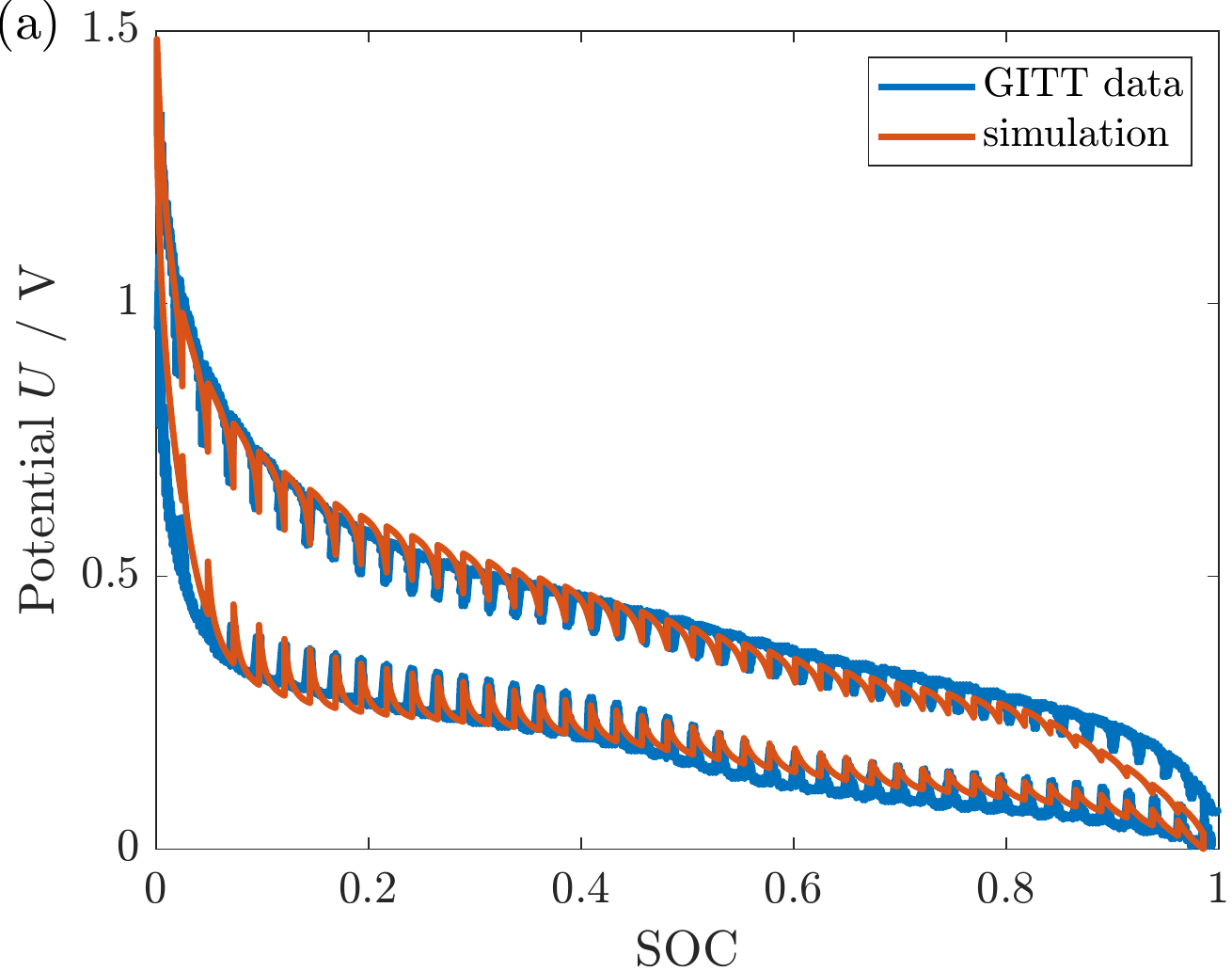}
	\hspace{0.5cm}
	\includegraphics[width=0.48\textwidth]{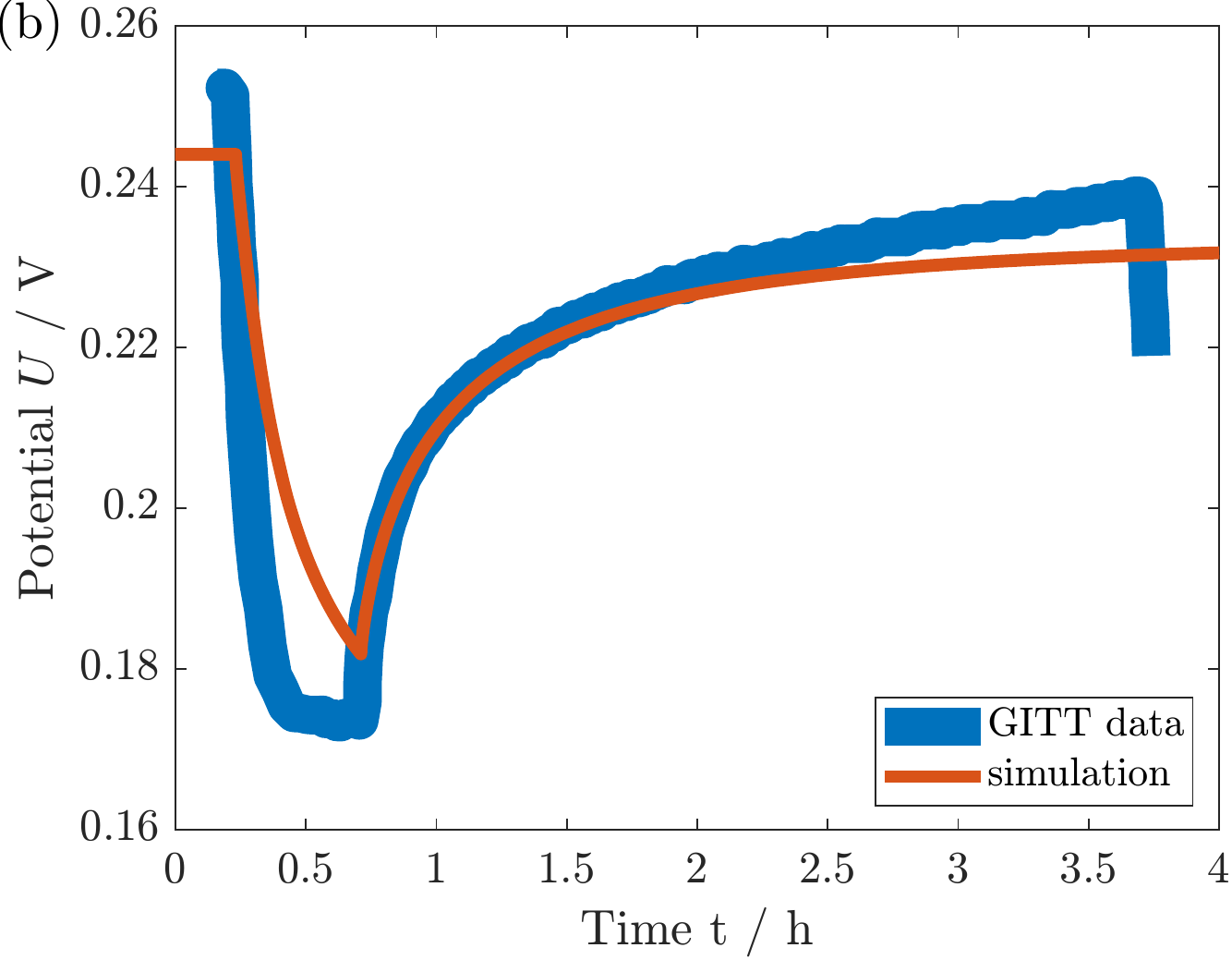}
	\caption{Impact of visco-elastoplastic SEI on the silicon potential.  (a) Simulation of the voltage hysteresis generated by a visco-elastoplastic SEI during a GITT procedure in comparison to the measurement \cite{Pan2019}. (b) Simulation of lithiation pulse and rest period in comparison to a single GITT pulse \cite{Pan2020}.}
	\label{fig:viscous-hysteresis}
    %\vspace{-0.5cm}
\end{figure*}

To investigate the influence of viscous SEI behavior on the voltage during slow cycling, we simulate the GITT measurement procedure with alternating $\mathrm{C}/20$ current steps and relaxation periods. We show the simulated voltage together with the voltage measured during the GITT experiment \cite{Pan2019} in \cref{fig:viscous-hysteresis}(a). The simulated voltage profile shows a good agreement with the experiment. The fitted viscosity value in \cref{fig:viscous-hysteresis}(a) is $\eta_\mathrm{SEI,0} = 15\,\mathrm{GPa\,s}^n$.
Only the potential difference between the two observed hysteresis phenomena stays approximately constant in the experiment but slightly decreases with increasing SOC in the simulation. The declining strain rate of the SEI for higher SOC causes the observed decrease of the viscous contribution in the simulation. More pronounced shear thinning behavior could reduce the deviation between simulation and experiment.

In \cref{fig:viscous-hysteresis}(b), we depict the simulation of a single lithiation pulse with a subsequent rest period in comparison to an experimental GITT pulse \cite{Pan2020}. The voltage curve demonstrates the transition between the lithiation potential and the measured GITT data points and vice versa. During the lithiation step, the simulation exhibits a slower decay from the rest potential to the lithiation potential compared to the experiment. Therefore, the timescale of the simulated lithiation step with the applied parameters is larger than the experimental value. For the subsequent rest period, our simulation and the experiment reveal similar timescales.
However, the experiment reveals an additional longer timescale as the voltage is not constant after three hours of relaxation.
A more sophisticated rheological model describing the visco-elastoplastic behavior could improve the timescales by the cost of additional parameters. Further, we will evaluate the influence of Young's modulus on the simulated timescales of GITT pulses in section \ref{sec:parameter-study}.

In the supporting information, we compare shear-thinning viscosity with Newtonian behavior. As illustrated in Fig. S6(a), Newtonian behavior leads to a more pronounced decrease of the viscous contribution to the voltage hysteresis with increasing SOC, which disagrees with the experiment. Therefore, we consider the power-law shear thinning model in our studies. However, due to the constant value of the viscosity, we can compare the Newtonian viscosity to the range of values presented in section \ref{sec:material-parameters}. The value $\eta_\mathrm{SEI} = 1.25\times 10^{14}\,\mathrm{Pa\, s}$ found for the Newtonian viscosity agrees with the range of $10^7\,\mathrm{Pa\, s} < \eta_\mathrm{SEI} < 10^{15}\,\mathrm{Pa\, s}$ discussed as reasonable in section \ref{sec:material-parameters}. The accordance indicates the suitability of considering viscous behavior to explain the amplified hysteresis observed during cycling.
Furthermore, the high fitting value of the viscosity matches the assumption of a stiff inner SEI layer.

Regarding the timescales, Newtonian viscosity and power-law shear-thinning viscosity reveal similar transition times during the GITT pulse depicted in S6(b). Thus, the choice of shear-thinning or Newtonian behavior does not significantly influence the transition profile between the different potential curves.

\subsection{Variation of Young's modulus}
\label{sec:parameter-study}

\begin{figure*}[htp]
	% Figure #?
	% width = 2 columns
	\centering
	\includegraphics[width=0.48\textwidth]{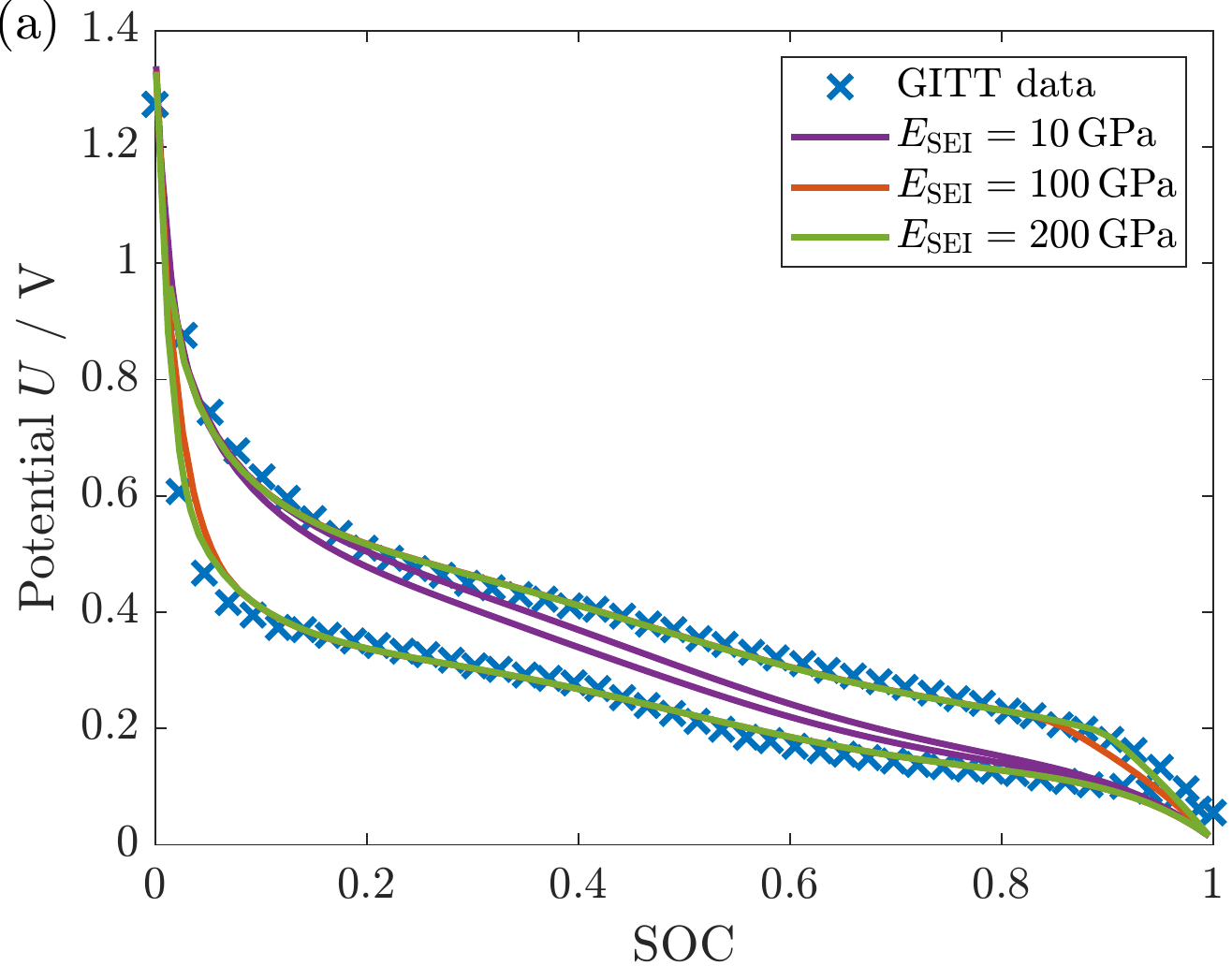}
	\hspace{0.5cm}
	\includegraphics[width=0.48\textwidth]{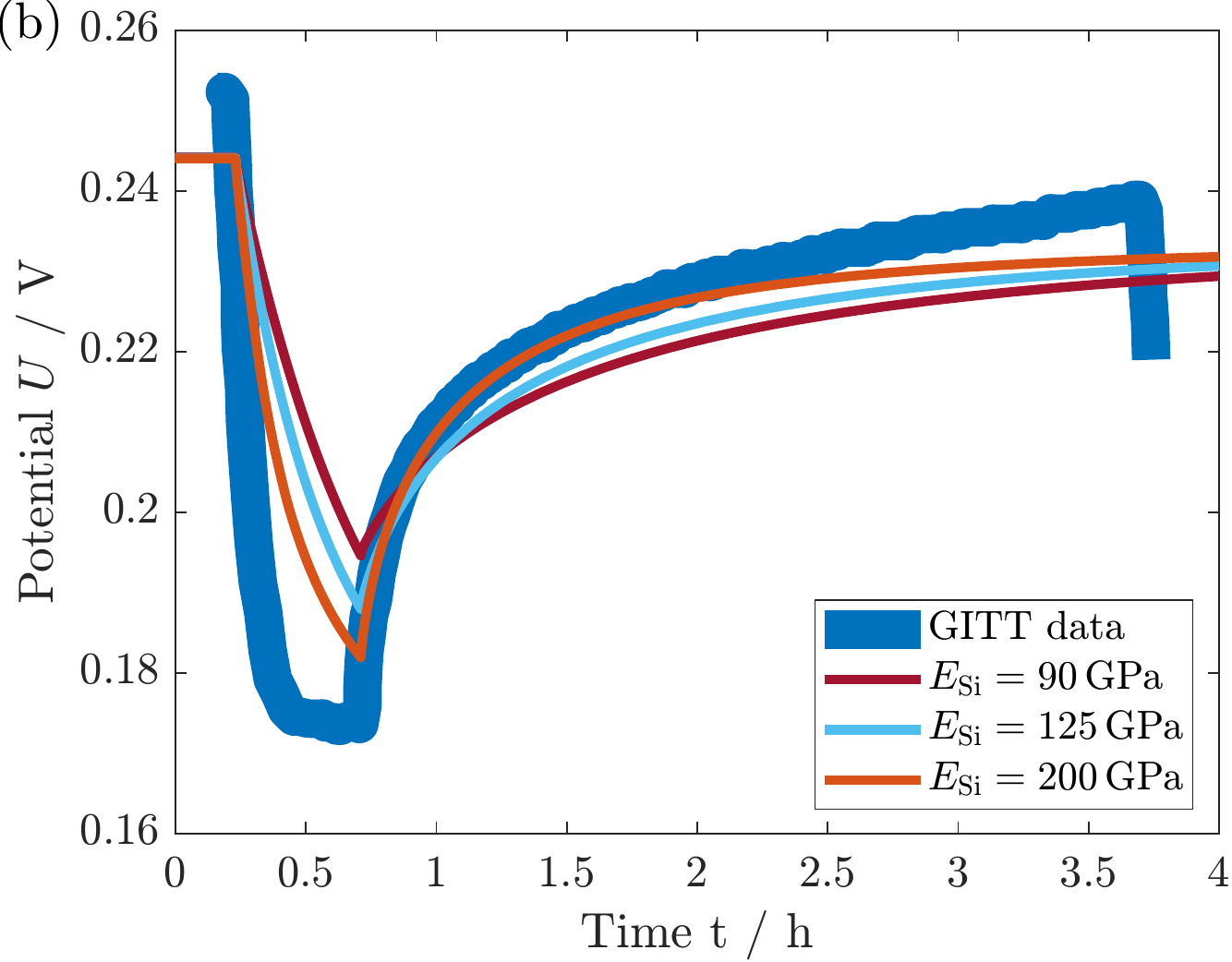}
	\caption{Effect of Young's modulus of the SEI and the silicon nanoparticle on the voltage hysteresis. (a) Impact of Young's modulus of the SEI on the shape of the silicon voltage hysteresis in comparison to experimental data \cite{Pan2019}. (b) Influence of Young's modulus of the silicon nanoparticle on the transition voltage profile in comparison to a single GITT pulse \cite{Pan2020}.}
	\label{fig:parameter-study}
    %\vspace{-0.2cm}
\end{figure*}

As the exact mechanical parameters of the SEI are ambiguous in experiments, we provide a parameter study to estimate the influence of Young's modulus $E_\mathrm{SEI}$. In \cref{fig:parameter-study}(a), we compare the open-circuit voltage hysteresis simulated for three different values of Young's modulus. For a small Young's modulus of the SEI, the simulated voltage hysteresis is small, and its shape does not match the experiment. In contrast, there is a reasonable agreement between simulation and experiment for medium and high values. The value $E_\mathrm{SEI} = 200\,\mathrm{GPa}$ shows the best agreement. This indicates that Young's modulus of the SEI is often underestimated and the nanoscale effect discussed in section \ref{sec:material-parameters} induces a further increase.

In the supporting information, Fig. S7(a) illustrates the timescale of a single lithiation pulse and relaxation for the medium and high parameter values of $E_\mathrm{SEI}$. Overall, the simulated timescales are comparable with the ones observed in the experiment. Increasing Young's modulus of the SEI leads to a slightly faster transition. However, the influence of Young's modulus of the SEI on the revealed timescale is essentially negligible.

Therefore, we analyze the mechanical properties of the silicon nanoparticle and its impact on the transition times. Due to the increase of Young's modulus reported at the nanoscale \cite{Chen2006, Mathur2007, Agrawal2008, He2008, Chen2015}, we vary Young's modulus of the silicon nanoparticle for constant Young's modulus of the SEI.
We compare the literature values of Young's modulus for bulk amorphous silicon of $90-125\,\mathrm{GPa}$ \cite{Kluge1988, Shenoy2010, Tan1972, Szabadi1998} with the modulus for silicon estimated at the nanoscale of $200\,\mathrm{GPa}$ from Refs. \cite{NasrEsfahani2019, Tang2012}. 
In \cref{fig:parameter-study}(b), we depict the voltage of a simulated GITT pulse for the variation of Young's modulus of silicon.
The highest value $E_\mathrm{Si}=200\,\mathrm{GPa}$ shows the best agreement with the experiment for the transition profile. The voltage relaxation for the lower values of $E_\mathrm{Si}$ is slightly too slow. Nevertheless, the lower values are still able to qualitatively describe the GITT data.
The variation reveals that a value for Young's modulus of the silicon nanoparticle, which is approximately a factor two higher than the literature bulk value, can best reproduce the timescale of the experiment. This indicates the importance of considering nanoscale effects for the mechanical description.

At the same time, we investigate the influence of Young's modulus of the silicon nanoparticle on the shape and size of the voltage hysteresis in the supporting information. Fig. S7(b) reveals that Young's modulus of silicon has only a minor influence on the voltage hysteresis in our model.

Therefore, the parameter study shows that high values of Young's modulus of the silicon nanoparticle and the SEI achieve the best agreement between the chemo-mechanical simulation and experimental data. This indicates the importance of nanoscale effects on the silicon voltage hysteresis.

\vspace{-0.2cm}
\section{Conclusions}

To conclude, the slow diffusion process or particle constriction inside electrodes can not cause the voltage hysteresis observed for silicon nanoparticle anodes. However, the expansion of silicon nanoparticles leads to an impact of the SEI or the silicon oxide layer on the stress and potential inside the anode. We propose a visco-elastoplastic silicon-SEI model, which can for the first time explain the open-circuit voltage hysteresis of a silicon nanoparticle anode measured with GITT. The plasticity of the SEI leads to different stresses inside the particle during lithiation and delithiation, inducing the observed voltage hysteresis. In addition, our SEI model qualitatively reproduces the volume hysteresis of silicon anodes. Furthermore, the viscous SEI behavior describes the difference between the voltage hysteresis observed for cycling at low currents and the open-circuit voltage from GITT after a relaxation period.
A variation of Young's modulus of the SEI reveals its crucial influence on the shape of the voltage hysteresis. 
The results indicate a stiff inner SEI layer.
Consequently, our work suggests that a soft SEI will mitigate the voltage hysteresis of silicon nano-anodes.
%\ \\\

\vspace{-0.3cm}
\begin{acknowledgments}
Lukas Köbbing gratefully acknowledges funding and support by the German Research Foundation (DFG) within the research training group SiMET under project number 281041241/GRK2218.
\end{acknowledgments}

\bibliography{refs}% Produces the bibliography via BibTeX.

\end{document}